# Landau level splitting in $Cd_3As_2$ under high magnetic fields

Junzhi Cao[1,2]†, Sihang Liang[1,2]†, Cheng Zhang[1,2]†, Yanwen Liu[1,2], Junwei Huang[3], Zhao Jin[3], Zhi-Gang Chen[4], Zhijun Wang[5], Qisi Wang[1,2], Jun Zhao[1,2], Shiyan Li[1,2], Xi Dai[5], Jin Zou[4,6], Zhengcai Xia[3], Liang Li[3], Faxian Xiu[1,2]*

[1] State Key Laboratory of Surface Physics and Department of Physics, Fudan University, Shanghai 200433, China

[2] Collaborative Innovation Center of Advanced Microstructures, Fudan University, Shanghai 200433, China

[3] Wuhan National High Magnetic Field Center, Huazhong University of Science and Technology, Wuhan 430074, China

[4] Materials Engineering, The University of Queensland, Brisbane QLD 4072, Australia

[5] Beijing National Laboratory for Condensed Matter Physics and Institute of Physics, Chinese Academy of Sciences, Beijing 100190, China

[6] Centre for Microscopy and Microanalysis, The University of Queensland, Brisbane QLD 4072, Australia

†These authors contributed equally to this work.

*Correspondence and requests for materials should be addressed to F. X. (E-mail: faxian@fudan.edu.cn)

**Abstract**

Three-dimensional topological Dirac semimetals (TDSs) are a new kind of Dirac materials that exhibit linear energy dispersion in the bulk and can be viewed as three-dimensional graphene. It has been proposed that TDSs can be driven to other exotic phases like Weyl semimetals, topological insulators and topological superconductors by breaking certain symmetries. Here we report the first transport experiment on Landau level splitting in TDS $Cd_3As_2$ single crystals under high magnetic fields, suggesting the removal of spin degeneracy by breaking time reversal symmetry. The detected Berry phase develops an evident angular dependence and possesses a crossover from nontrivial to trivial state under high magnetic fields, a strong hint for a fierce competition between the orbit-coupled field strength and the field-generated mass term. Our results unveil the important role of symmetry breaking in TDSs and further demonstrate a feasible path to generate a Weyl semimetal phase by breaking time reversal symmetry.

**Introduction**

The peculiar band structure of graphene makes it a text-book Dirac material and a promising candidate for next-generation electronic devices.[1, 2, 3] It has been found that Dirac fermions with linear band dispersion can give rise to various physical phenomena, such as quantum Hall effect, Andreev reflection and Klein tunneling.[1, 2, 4, 5, 6] Driven by the excellent properties of Dirac materials, topological Dirac semimetals (TDSs), adopting a similar band structure to graphene but in the bulk form, have been theoretically proposed in several systems, including β-$BiO_2$, $Na_3Bi$ and $Cd_3As_2$.[7, 8, 9] In a TDS, the conduction band and the valence band contact each other only at some discrete points (Dirac nodes) in the momentum space. These Dirac nodes are degenerated and they consist of several overlapping Weyl nodes with opposite chirality in the presence of time reversal symmetry (TRS) and inversion symmetry (IS).[7, 8] In the meantime, the additional crystalline point-group symmetry is required to preserve the overlapping Weyl nodes from annihilation in TDSs.[8] Therefore, the three-dimensional (3D) Dirac nodes always occur along the high symmetry directions in the momentum space.

One of the most striking features of TDSs is the presence of various exotic phases, like Weyl semimetals, topological insulators and topological superconductors, by breaking certain symmetries in the system.[8] It has been theoretically predicted that breaking TRS or IS can remove the degeneracy of the Dirac nodes,[7] resulting in a Weyl semimetal phase with opposite chiral Weyl node pairs.[10] Such an emerging phase promises many intriguing transport phenomena, such as chiral magnetic effect[11] and nonlocal transport,[12, 13] thus developing a possible basis for new electronic applications like chiral battery or quantum amplifier.[11]

Soon after the theoretical predictions, extensive experimental efforts have been devoted to the discovery of the TDS phase in the representative materials, $Na_3Bi$[14] and $Cd_3As_2$.[15, 16, 17, 18] Photoemission spectroscopy unveils a pair of 3D Dirac nodes in $Cd_3As_2$ locating on the opposite sides of the Brillouin zone center (Γ point) which are protected by the crystal symmetry.[17, 19] Transport measurements reveal an ultra-high mobility, a giant linear magnetoresistance and a nontrivial Berry phase owing to the linear band dispersion and concomitant Dirac fermions.[20, 21, 22, 23] Field-induced Landau level splitting in $Cd_3As_2$ has been observed by scanning tunneling microscopy,[15] where a perpendicular field incurs the doublet structure of the Landau levels. However, to date, the transport experiments were mostly performed in a low magnetic field regime (<15 T) and thus unable to track the Landau level splitting in the quantum limit. In addition, a well-controlled field direction to avoid the possible crystal symmetry breaking is a prerequisite for realizing the Weyl semimetal phase, which could be accessible in the transport experiments.

In this study, we report the low-temperature magnetotransport properties of $Cd_3As_2$ single crystals under high magnetic fields. Shubnikov-de Haas (SdH) oscillations clearly resolve strong Landau level splitting at pulsed magnetic fields. The spacing of the split Landau levels, defined as the spatial difference of the split peaks, changes with the field direction, revealing a combination of the orbital and Zeeman splitting. Significantly, we observed an angular dependent Berry phase at high magnetic fields, a signature of the competition between the orbit-coupled field strength and the generated mass term. These findings serve as the evidence for the isolation of Weyl nodes and the emerging Weyl fermions in $Cd_3As_2$ by breaking the TRS.

**Results**

**Structure characterizations.** Transmission electron microscopy (TEM) was carried out to determine the structural characteristics of the synthesized $Cd_3As_2$ crystals. Figure 1a shows a high-resolution TEM image of a $Cd_3As_2$ thin flake on a holey carbon grid, revealing a perfect antifluorite ($M_2X$)-type crystalline structure. The inset is a low magnification TEM image of the examined flake. Figure 1b shows a typical energy dispersive x-ray spectrum (EDX) with an atomic ratio of Cd:As=3:2. Consistent with the TEM results, the x-ray diffraction (XRD) peaks can be indexed as series of {112} planes, which verifies the high crystallinity of the single crystals (Figure 1c). The crystal structure of our $Cd_3As_2$ samples is found to be $I4_{1/acd}$. Its unit cell is tetragonal with $a$=12.633(3) Å and $c$=25.427(7) Å. Each unit cell contains 96 Cd atoms and 64 As atoms.

**Hall effect measurements.** A Hall bar device with a standard six-terminal geometry was fabricated for the transport measurements, as schematically illustrated in Figure 1d. A constant current was applied within the {112} atomic planes while the magnetic field was titled from perpendicular to parallel to the {112} planes, as depicted by the blue arrow. Figure 1e provides the temperature dependence of longitudinal resistivity $\rho_{xx}$ at zero magnetic field. The $\rho_{xx}$-$T$ curve describes a typical metallic behavior of $Cd_3As_2$ due to the semimetal band structure. One of the most fascinating features of $Cd_3As_2$ is the ultra-high mobility deriving from the linear band dispersion. A typical $Cd_3As_2$ sample yields a high electron mobility of $\mu = 2 \times 10^4$ cm$^2$ V$^{-1}$ s$^{-1}$ at room temperature from the Hall effect measurements, in agreement with previous studies.[22] In fact, most of our samples (refer to Supplementary Figure 1~4 and Supplementary Table 1) have a room-temperature mobility in the range of $(1\sim5) \times 10^4$ cm$^2$ V$^{-1}$ s$^{-1}$. Figure 1f shows the temperature dependence of mobility (red curve), which dramatically increases to $1.9 \times 10^5$ cm$^2$ V$^{-1}$ s$^{-1}$ at 2.6 K. The significant improvement of the mobility can be attributed to the alleviated phonon scattering at very low temperatures. A previous study revealed a wide-range distribution of resistivity at low temperatures in $Cd_3As_2$, corresponding to different electron mobility, which is extremely sensitive to disorder.[21] The residual resistivity of our sample (20 μΩ cm) and Hall mobility (1.9×10$^5$ cm$^2$ V$^{-1}$ s$^{-1}$) agree with their trend. The carrier density of the sample also exhibits a negligible change with temperature and reaches a relative low value of $n_e = 1.67 \times 10^{18}$ cm$^{-3}$ at 2.6 K (Figure 1f, blue curve). Such a low carrier density makes it easier for the Fermi level to reach low Landau levels.[24]

**Fermi surface and quantum oscillation analysis.** In order to probe the Fermi surface (FS) of $Cd_3As_2$, we carried out the magnetotransport measurements using a physical properties measurement system (PPMS, up to 9 T). Figure 2a and 2b depict the magnetoresistivity (MR) of $Cd_3As_2$ with a parabolic and a quasi-linear behavior near the zero field and at large fields (*B*>4 T), respectively. The parabolic behavior is originated from the orbit contribution and it becomes more pronounced at high temperatures,[21] while the quasi-linear part survives up to 375 K at large fields although the MR ratio drops from 32,500 % at 2.6 K to 600 % at 375 K. The reduction of the MR ratio is believed to be associated with the temperature-sensitive phonon scattering.[8]

Apart from the giant MR, more than 90 % of our samples exhibit strong quantum oscillations which are attributed to the Dirac band structure and the resultant ultra-high mobility. Evident SdH oscillations can be well-resolved when the temperature is below 30 K in both longitudinal MR and the Hall signal, as shown in Figure 2b and 2c, respectively, where the oscillations can be tracked down to 3 T. The tilting angle $\theta$ in Figure 2a is defined as the angle between the magnetic field $B$ and the normal direction of the (112) plane (also refer to Figure 1d). It is found that the rotation of the sample from perpendicular to parallel to the crystal plane causes the decrease of the oscillation amplitude, but the oscillation frequency remains nearly unchanged (<5%, refer to Supplementary Figure 1b and Supplementary Note 1). The origin of the angle-dependent MR ratio and oscillation amplitude will be discussed later.

To fundamentally understand the SdH oscillations, we calculate the oscillation frequency ($F$) to be 61.8 T, corresponding to a periodicity of $\Delta(1/B) = 0.0162 \text{ T}^{-1}$. From the equation $F = (\phi_0/2\pi^2)S_F$ with $\phi_0 = h/2e$,[25] the cross-section area of the FS can be determined as $S_F = 5.89 \times 10^{-3} \text{ Å}^{-2}$. Despite the strong angular dependence of the amplitude, the nearly unchanged oscillation periodicity suggests a negligible anisotropy of the Fermi sphere (Supplementary Figure 1b). Thus, by assuming a circular cross-section, the Fermi vector of $k_F = 0.043 \text{ Å}^{-1}$ can be extracted. The SdH amplitude as a function of temperature can be analyzed to extract key parameters of the carrier transport. The temperature-dependent amplitude $\Delta\sigma_{xx}$ is described by $\Delta\sigma_{xx}(T)/\Delta\sigma_{xx}(0) = \lambda(T)/\sinh(\lambda(T))$, and the thermal factor is given by $\lambda(T) = 2\pi^2 k_B T m_{cycl}/(\hbar eB)$, where $k_B$ is the Boltzmann's constant, $\hbar$ is the reduced plank constant, and $m_{cyc} = E_F/v_F^2$ is the cyclotron mass.[26, 27] By taking conductivity oscillation amplitude and performing the best fit to the $\Delta\sigma_{xx}(T)/\Delta\sigma_{xx}(0)$ equation, $m_{cyc}$ is extracted to be $0.05\ m_e$. Using the equation $v_F = \hbar k_F/m_{cyc}$, we can obtain the Fermi velocity of $v_F = 1.00 \times 10^6\ m\ s^{-1}$ and the Fermi energy of $E_F = 286$ meV. A long mean free path of $\tau = 101$ nm can be estimated from the Dingle plot as shown in Figure 2e. Table 1 summarizes the estimated parameters derived from the SdH oscillations when $B \leq 9$ T.

**Magnetotransport under high magnetic fields.** To search the possible new phases involving symmetry breaking, it is necessary to apply higher magnetic field to reach the lower Landau levels. Indeed, a pulsed field of 52 T drives the sample to the 2$^{nd}$ Landau level, as shown in Figure 3a, where the longitudinal resistivity is plotted against the magnetic field (up to 52 T). The angular and temperature dependence of high field longitudinal resistivity along with the related Landau fan diagrams are plotted in Supplementary Figure 5~7. Here, we use integers to denote peaks and half integers to represent valleys,[25] from which the splitting of the 2$^{nd}$ and the 3$^{rd}$ Landau levels can be clearly witnessed. The Landau level splitting observed here is considered to be the joint effect of the Zeeman and the orbit contributions.[15] It has been theoretically predicted that, under a relatively large magnetic field, the FS topology as well as the topological charge enclosed by the FS can be largely tuned by varying the field strength and direction.[9] Thus the TDS system shows a variety of distinct topological phase transitions driven by breaking symmetries.[9, 28] When a magnetic field is applied, the TRS in the system is no longer preserved.[9, 24] By considering the exchange couplings induced by the external field, we can in general separate the field-dependent Hamiltonian to the orbital-dependent part and the orbital-independent part as $H_{ex1} = h_1 \sigma_z \otimes \tau_z$ and $H_{ex2} = h_2 \sigma_z \otimes I$, respectively,[9, 15] where $h_1$ and $h_2$ are the field strength along the z direction, $\sigma_z$ and $\tau_z$ are Pauli matrices for spin and pseudospin, respectively. If the field only couples to spin ($H_{ex1} = 0$), the FS will split into two concentric spheres. If the field couples to spin and orbit both ($H_{ex1} \neq 0$), the FS will split into two separate Weyl pockets.[9, 21] Furthermore, if the perturbation of the field on the crystal symmetry is considered, a mass term accompanied by the gap opening will be introduced to the Weyl nodes. Thus there will be a competition between the orbit-coupled field strength $h_1$ and the field-generated mass term *m*. When $h_1$ is larger than *m*, the Weyl semimetal phase can be developed.[9]

In order to clarify the respective contributions of the orbit and the Zeeman term to the Landau level splitting, we performed the angle-dependent magnetotransport measurements under high fields. In Figure 3b, when tilting the sample from *θ*=8º to 78.5º, the splitting features remain well-resolved and the split spacing changes with *θ*. Theoretically, it was predicted that the orbital-dependent splitting is highly sensitive to the field direction while the Zeeman term shows no significant angular dependence when the field strength is fixed.[15] Also, the orbital-dependent splitting can reach the maximum when the field is along the [001] direction and vanishes when the field is perpendicular to the [001] direction.[15] In our experiments (Figure 3b), the split spacing for the 2.5$^{th}$ peak shows less angular dependence than that of the 3$^{th}$ and the 3.5$^{th}$ peaks. Thus, the Zeeman term presumably dominates in the 2.5$^{th}$ peak splitting while for the 3$^{th}$ and the 3.5$^{th}$ peaks the orbital term has a large contribution. The Landau level splitting in another sample also shows a strong angular dependence as shown in Supplementary Figure 8.

Figure 3c displays the longitudinal magnetoresistivity oscillations under high magnetic fields at different temperatures (refer to Supplementary Figure 7 for the original data). The splitting of the Landau levels becomes less resolved with increasing the temperature. Owing to the large effective Landé factor $g_s^*$ in Cd$_3$As$_2$ ($g_s^* = 40$),[15, 22] the obtained Zeeman splitting energy ($g_s^* \mu_B B$) should be considerably large, especially in the high magnetic fields (corresponding to the lower bands).[24] When raising the temperature, the thermal energy $k_B T$ of electrons becomes larger, and in some materials it turns to be comparable with the Zeeman splitting energy. Between 4.2 and 80 K, the splitting of the lowest two levels remains evident primarily due to the fact that the splitting energy (from Zeeman and orbit) at high fields is still larger than the thermal energy $k_B T$.

In a Dirac system, there exists a "zero mode" that does not shift with the field, leading to a nontrivial $\pi$ Berry phase.[24, 29, 30] According to the Lifshitz-Onsager quantization rule: $S_F \frac{\hbar}{eB} = 2\pi \left( n + \frac{1}{2} - \frac{\phi_B}{2\pi} \right) = 2\pi (n+\gamma)$,[29] the offset $\gamma$ in the Landau fan diagram gives the Berry phase $\phi_B$ by $\gamma = \frac{1}{2} - \frac{\phi_B}{2\pi}$, where $\hbar$ is the reduced Planck's constant and $e$ is the elementary charge. For a nontrivial Berry phase, $\gamma$ should be 0 or 1. However, the exchange couplings with the magnetic field (including the Zeeman and the orbital terms) lift the degeneracy at the Dirac point. As aforementioned, the field will affect the crystal symmetry if the direction is not along the [001] direction. In this scenario, a gap will emerge as long as the orbit-coupled field strength is smaller than the mass term, resulting in the shift of the Berry phase. Consequently, a change in the Berry phase can be expected at higher fields.[24] Here, we extrapolate the value of the offset $\gamma$ in the field regimes of 5-7 T (corresponding to 9th to 11th Landau levels), 7-10 T (7th to 9th Landau levels), and 10-15 T (5th to 7th Landau levels). The reason we choose the regime of $B \leq 15$ T to perform the Berry phase fitting is that at high fields the Landau fan diagram itself becomes nonlinear and inevitably introduces a large deviation in the linear fitting process (see Supplementary Figure 6 and Supplementary Note 2 for more details). From the offset, the Berry phase $\phi_B$ can be acquired via the Lifshitz-Onsager equation, as summarized in Figure 4, where $\phi_B$ gradually develops an angular dependence at high field regimes. In comparison, the corresponding Berry phase shows no such dependence at the low field regime (5-7 T). Taking $\theta=8°$ as an example, the Berry phase changes from $(0.67\pm0.05)\pi$ to $(0.27\pm0.06)\pi$ when the external field increases. Apparently, our experiments show a tendency that the Berry phase changes from nontrivial to trivial as the magnetic field is increased when $\theta$ is small.

**Discussion**

It has been demonstrated that an additional phase shift will arise from the curvature of the FS, changing from 0 for a quasi-2D cylindrical FS to $\pm\frac{\pi}{4}$ for a corrugated 3D

FS, with the precise value determined by the degree of two-dimensionality.[25] However, from our results, the disparity of the Berry phase is much larger than $\frac{\pi}{4}$. And the Berry phase acquired from the low field regime does not have such a crossover from nontrivial phase to trivial state at different field directions. It suggests that the angular dependent Berry phase is related to a field-generated phase transition instead of the Fermi surface curvature. The Berry phase tends to be trivial at small tilting angles while it retains nontrivial at large angles (for example, $\theta$= 78.5°). No aperiodic behavior was observed at $\theta$= 78.5° in our field range (refer to Supplementary Figure 6). These evidences suggest that when $\theta$ is 78.5°, the system remains nontrivial states and no gap is induced even under high magnetic fields. It should be noted that the field is not applied along the [001] direction when $\theta$= 78.5°, in which the crystal symmetry is expected to be broken and a mass term could be in principle introduced to the system. Although we cannot fully exclude other possible mass-generation mechanisms in the system which could be induced by high magnetic fields, this novel dependence of Berry phase on field direction and strength matches the phase diagram proposed in previous theoretical study.[9] Therefore, the presence of the nontrivial phase at $\theta$=78.5° in our experiments provide evidence that the field-generated mass term could be removed through the possible formation of a Weyl semimetal phase, consistent with the recent predictions.[7, 9, 31]

After clarifying the field effect on the $Cd_3As_2$ crystal, we revisit the observed linear MR. According to the recent study, the giant linear MR in TDSs may result from the collapse of a novel mechanism from backscattering in the presence of the magnetic field.[21] As discussed previously, the magnetic field can break the TRS and affect the crystal symmetry simultaneously.[9] The crystal symmetry breaking will generate a gap in the Dirac nodes.[9, 15] By controlling the field direction, we can induce discrete Weyl nodes to retain the gapless feature and nontrivial Berry phase. However, as long as the orbital-dependent splitting is strong enough to eliminate the induced gap, a Weyl semimetal phase can still emerge and the protection from backscattering survives.[9] Owing to the field-direction sensitive nature of the orbital-dependent splitting, the MR ratio and the oscillation amplitude are closely related to the field direction as well.

In conclusion, we observe the Landau level splitting and an altered Berry phase under the high magnetic fields in the ultra-high mobility $Cd_3As_2$ single crystals. The orbital-dependent splitting and the Berry phase can be significantly affected by the direction of the applied field. Our study demonstrates the possibility of inducing Weyl semimetal phase in TDSs by breaking symmetries. Further improvement could be accomplished by using local magnetic dopants to achieve an intrinsic Weyl semimetal. After submitting the manuscript, we became aware of the related studies reporting the observation of an intrinsic Weyl semimetal in TaAs class.[32, 33, 34, 35]

**Methods**

**Single crystal growth.** High-quality $Cd_3As_2$ single crystals were synthesized by self-flux growth method in a tube furnace. Stoichiometric amounts of high-purity Cd powder (4N) and As powder (5N) elements were placed inside an alumina crucible. The molar ratio of Cd and As was 8:3. After mixing two elements uniformly, the alumina crucible was sealed in an iron crucible under argon atmosphere. The iron crucible was heated to 800-900 °C and kept for 24 hours, then slowly cooled down to 450 °C at 6 °C per hour. After that, the crucible was kept at 450 °C for more than one day then cooled naturally to room temperature. The superfluous Cd flux was removed by centrifuging in a vacuum quartz tube at 450 °C.

**Acknowledgements**

This work was supported by the National Young 1000 Talent Plan, Pujiang Talent Plan in Shanghai, National Natural Science Foundation of China (61322407, 11474058), and the Chinese National Science Fund for Talent Training in Basic Science (J1103204). J.C. and S.L. appreciate Juan Jiang, Lanpo He and Yao Shen for helps in material synthesis. F.X. thanks Yuanbo Zhang for helpful discussions. Part of the sample fabrication was performed at Fudan Nano-fabrication Laboratory.

**Author contributions**

F.X. conceived the ideas and supervised the overall research. J.C. and S.L. synthetized $Cd_3As_2$ single crystal with the help from Q.W. and J.Z. C.Z., Y.L., J.H., Z.J., Z.X. and L.L. performed the magnetotransport measurements. J.C., S.L. and C.Z. analyzed the transport data. Z-G.C. and J.Z. performed crystal structural analysis. Z.W., X.D. and S.L. provided suggestions and guidance of experiments. C.Z. and F.X. wrote the paper with helps from all other co-authors.

**Additional information**

Supplementary information is available in the online version of the paper. Reprints and permissions information is available online at www.nature.com/reprints. Correspondence and requests for materials should be addressed to F.X.

**Competing financial interests**

The authors declare no competing financial interests.

**Figure Captions**

**Figure 1| Structural and electrical properties of $Cd_3As_2$ bulk crystals. a,** A high-resolution TEM image of a $Cd_3As_2$ thin flake on a holey carbon grid, revealing a perfect crystalline structure. Inset is a low magnification TEM picture. The white and black scale bars correspond to 2 nm and 1 μm, respectively. **b,** A typical EDX spectrum showing the atomic ratio of Cd:As=3:2. **c,** X-ray diffraction patterns of the single crystal $Cd_3As_2$. The peak position shows that the sample surface is in (112) plane. **d,** A constant current was applied within the {112} atomic planes while the magnetic field was tilted in the x-z plane, as depicted by the blue arrow. **e,** The longitudinal resistivity $\rho_{xx}$ as a function of temperature, showing a typical metallic behavior. **f,** The temperature-dependent mobility and carrier density from 2.6 to 300 K. At 2.6 K, the mobility reaches $1.9 \times 10^5$ cm$^2$ V$^{-1}$ s$^{-1}$.

**Figure 2| Low magnetic field transport measurements ($B\leq9$ T). a,** The angular dependence of longitudinal resistivity $\rho_{xx}$ at 2.6 K. The SdH oscillations are observed at different angles. The amplitude of the oscillation decreases as the angle $\theta$ becomes larger. **b,** The longitudinal resistivity at different temperatures at $\theta=0°$. The critical temperature is found to be 30 K, above which the oscillation is not observable. **c,** The Hall signal $R_{xy}$ of the sample from 3 to 300 K. **d,** Normalized conductivity amplitude $\Delta\sigma_{xx}(T)/\Delta\sigma_{xx}(0)$ versus temperature. The outcome can be fitted with the equation $\Delta\sigma_{xx}(T)/\Delta\sigma_{xx}(0)=\lambda(T)/(\sinh(\lambda(T))$ and the R-square is higher than 0.999 (the coefficient of multiple determination). The error bars were estimated to be 5 % of the normalized oscillation amplitude. **e,** Dingle plots of $\log[\Delta R/R \cdot B\sinh\lambda]$ versus $1/B$ at $\theta=0°$.

**Figure 3| The splitting of Landau levels in the high magnetic field. a,** The longitudinal resistivity $\rho_{xx}$ at $\theta=8°$ at 4.2 K. Landau levels are labeled by different colors with the resistivity peaks being integers and the valleys being half-integers. **b,** The angular dependence of longitudinal resistivity at 4.2 K. The SdH oscillations are observed at different angles. **c,** The longitudinal resistivity data at different temperatures at $\theta=8°$. The SdH oscillations persist up to 80 K.

**Figure 4| The angular dependent Berry phase $\Phi_B$ at different magnetic field regimes.** The Berry phase was fitted in the regimes of **a**, 5-7 T (corresponding to 9[th] to 11[th] Landau levels); **b**, 7-

10 T (7$^{th}$ to 9$^{th}$ Landau levels); and **c**, 10-15 T (5$^{th}$-7$^{th}$ Landau levels). The Berry phase develops an angular dependence at high magnetic fields, suggesting a field-induced phase transition. The error bars were generated from the linear fitting process in the Landau-fan diagrams.

**Table 1| Estimated parameters from the SdH oscillations ($B$≤9 T).** Transport parameters including the effective mass $m^*$, Fermi surface $S_F$, Fermi vector $K_F$, carrier lifetime $t$, Fermi velocity $v_F$, mean free path $l$, and Fermi energy $E_F$, can be extracted from the SdH oscillations.

# Figure 1

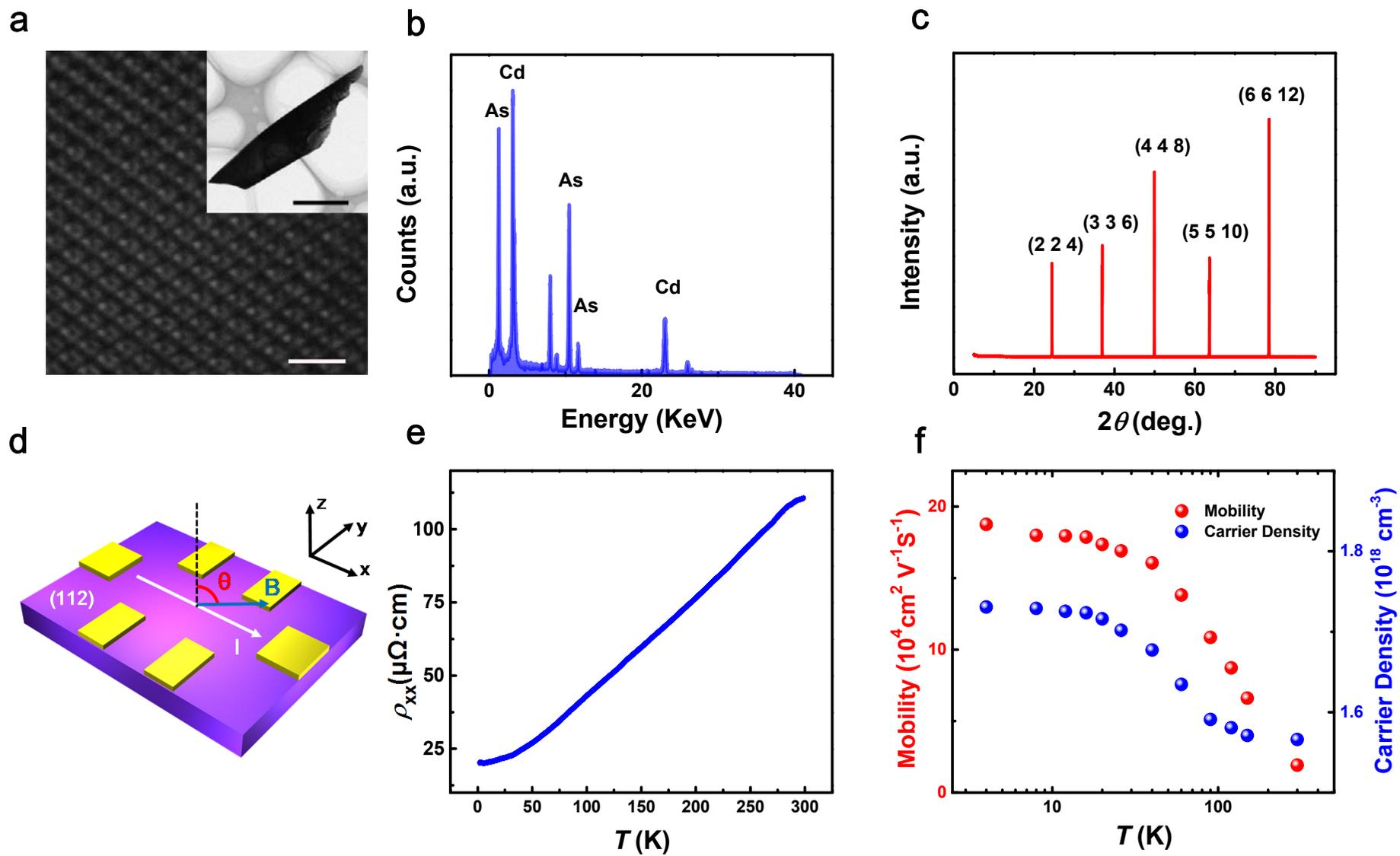

Figure 2

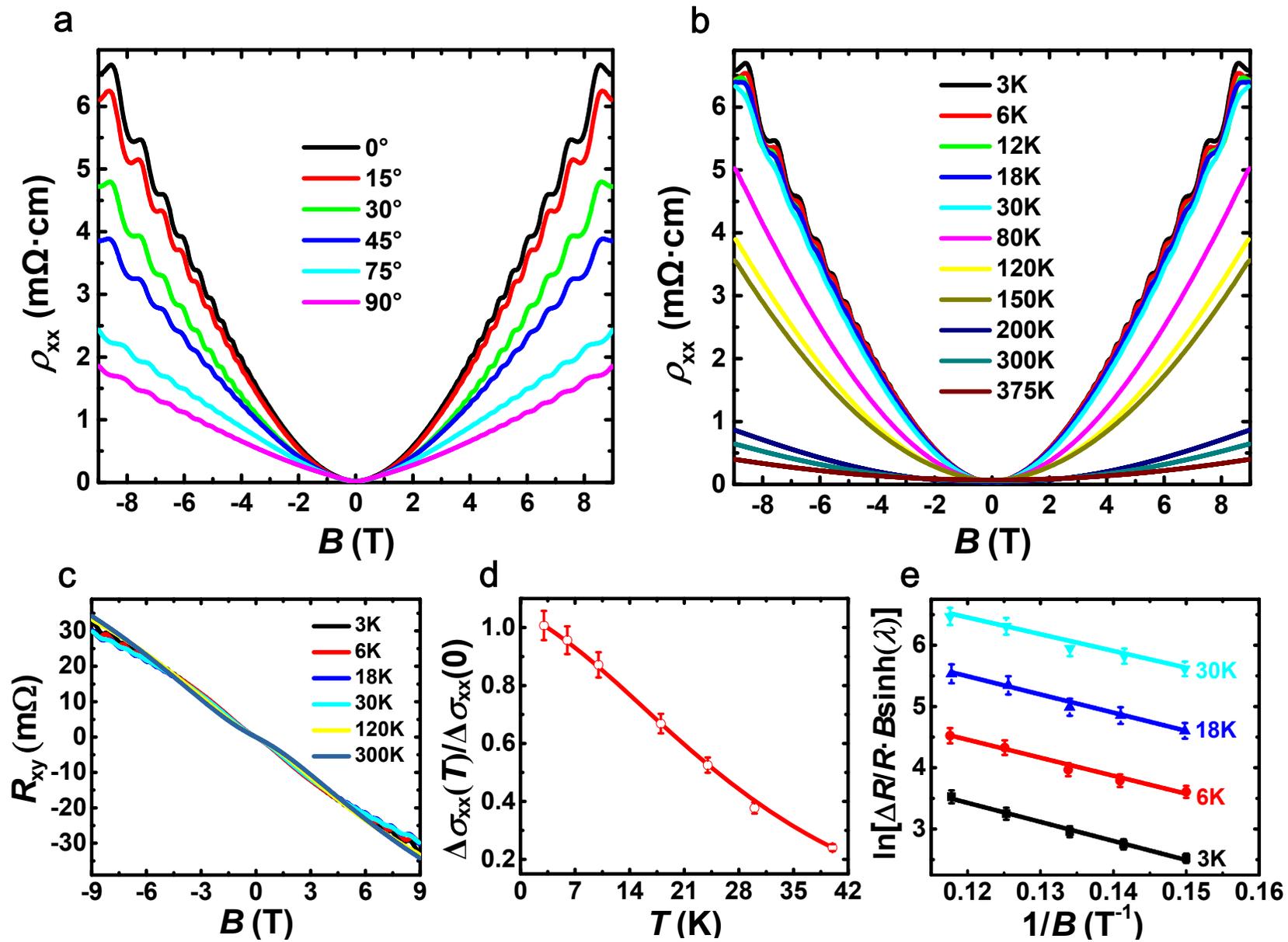

Figure 3

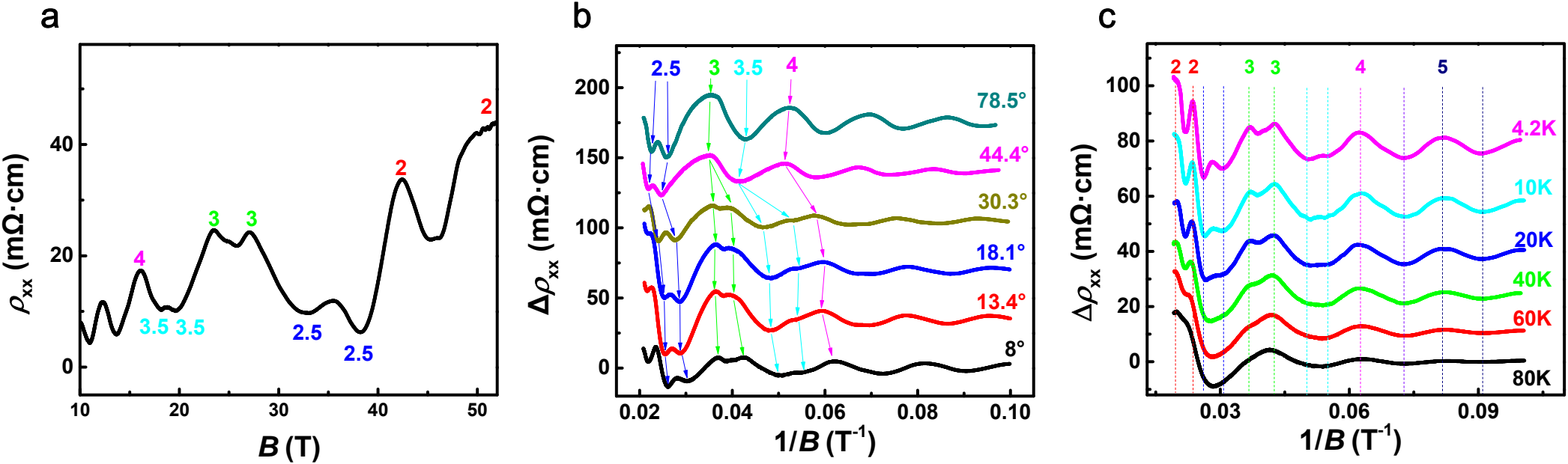

**Figure 4**

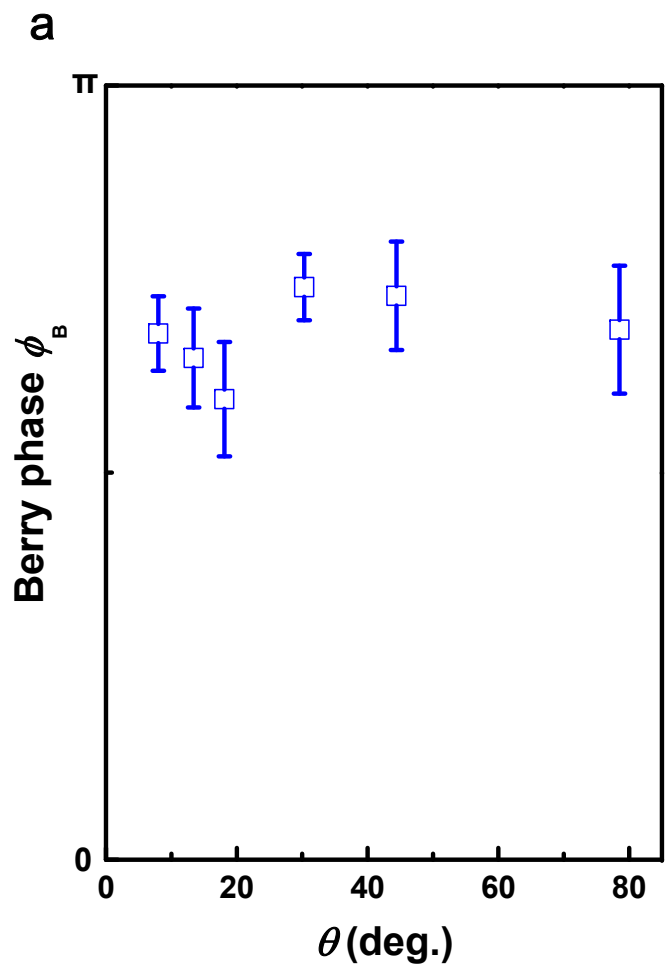 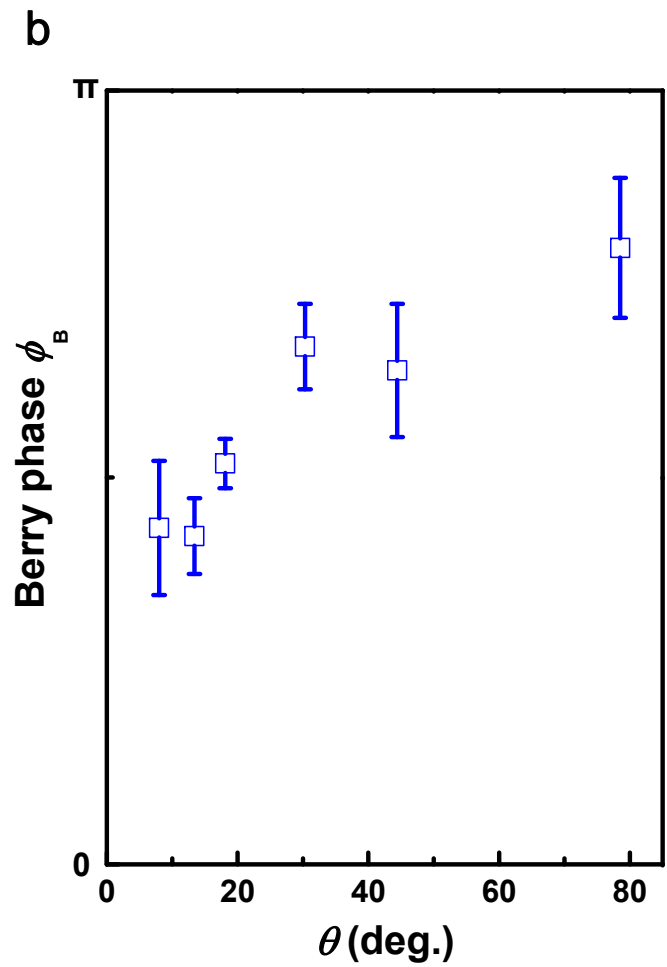 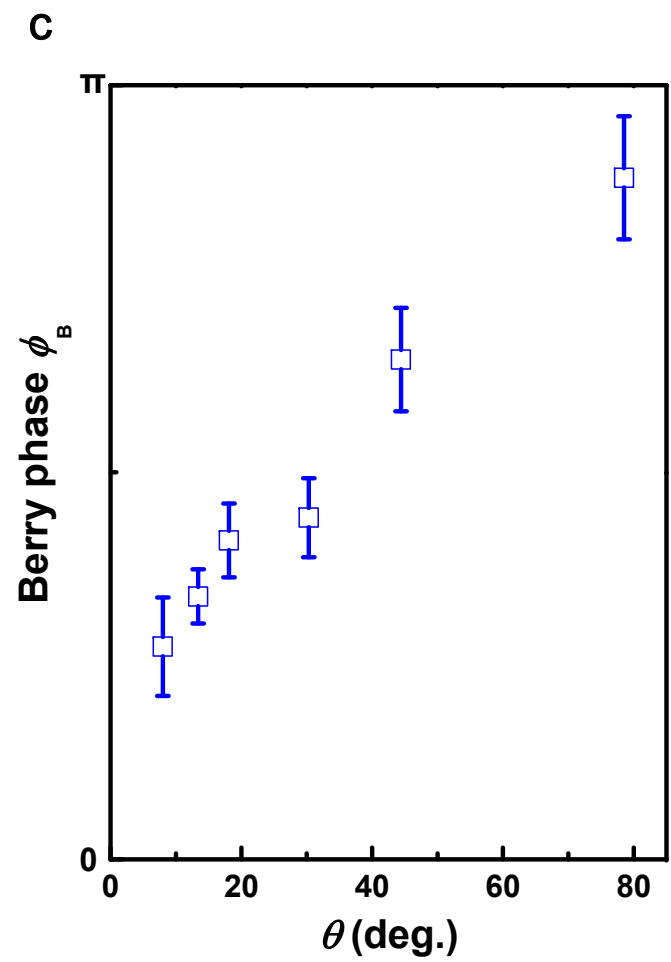

**Table 1**

| | $m^*$ | $S_F$ (Å$^{-2}$) | $k_F$ (Å$^{-1}$) | $t$ (s) | $v_F$ (m/s) | $l$ (nm) | $E_F$ (meV) |
|---|---|---|---|---|---|---|---|
| Sample 1 | $0.05m_0$ | $5.89\times10^{-3}$ | 0.043 | $1.41\times10^{-13}$ | $1.00\times10^6$ | 101 | 286 |